\documentclass[preprint,aps,prc,showpacs,nofootinbib]{revtex4}
\usepackage{amsmath}
\usepackage{amssymb}
\usepackage{graphics}
\usepackage{multirow}

\usepackage{epsfig}
\newcommand{\Li}{\mbox{Li}_2}

\newcommand\ba{\begin{eqnarray}}
\newcommand\ea{\end{eqnarray}}
\newcommand\be{\begin{equation}}
\newcommand\ee{\end{equation}}
\newcommand\nn{\nonumber}
\newcommand{\bas}{\begin{eqnarray*}}
\newcommand{\eas}{\end{eqnarray*}}

\newcommand{\TPE}{\ensuremath{2\gamma E\ }}
\begin{document}

\title{Updated analysis of recent results on electron and positron elastic scattering on proton}

\author{V.~V.~Bytev}
\email{bvv@jinr.ru}
\affiliation{\it JINR-BLTP, 141980 Dubna, Moscow region, Russian Federation}

\author{E.~Tomasi-Gustafsson}
\email{egle.tomasi@cea.fr}
\affiliation{\it DPhN, IRFU, CEA, Universit\'e Paris-Saclay, 91191 Gif-sur-Yvette Cedex, France}

%\date{\today}
%\pacs{25.30.Bf, 13.40.-f, 13.40.Gp}

\begin{abstract}
We discuss recent experimental results concerning the cross section ratio of  positron over electron elastic scattering on protons, and compare with the predictions of a pre-existent calculation. The deviation from unity of this ratio,  $i.e.$,  a charge asymmetry different from zero,  is the signature of contributions beyond the Born approximation. After reviewing the published results, we compare the elastic data to a calculation which includes the diagram corresponding to two-photon exchange. It turns out that all the data on the cross section ratio, in the limit of their precision, do not show evidence of enhanced two-photon contribution beyond the expected percent level. Our results confirm that experimental evidence for a large contribution of two-photon exchange is not yet found. 
%for which a relative contribution of $\simeq$ 6\%, with respect to the main (one photon exchange) term, would be necessary.
\end{abstract}

\maketitle
%%%%%%%%%%%%%%%%%%%%%%%%%%%%%%%%%
\section{Introduction}
%%%%%%%%%%%%%%%%%%%%%%%%%%%%%%%%%

Elastic electron-proton scattering is the object of large experimental and theoretical effort since many decades. Since the works that valued the Nobel Prize to R. Hofstadter in 1967, it is a privileged way to learn about the proton internal structure. Assuming that the interaction occurs through the exchange of one virtual photon with four-momentum $q^2<0$ ($q^2=-Q^2$) a simple and elegant formalism allows to express the proton electromagnetic current in terms of two electromagnetic - Pauli ($F_1$) and Dirac ($F_2$) - or, alternatively, the Sachs form factors  (FFs): $G_E$ and $G_M$. The experimental observables as the cross section and the polarization observables allow to  directly access these quantities (for a review, see \cite{Pacetti:2015iqa}).

In recent years large experimental and theoretical work is devoted to this subject due to
the possibility of very precise measurements at large transferred  momentum. The development of 100\% duty cycle electron machines as Jefferson Lab (JLab), with highly polarized electron beams, the construction of large solid angle spectrometers and detectors, the development of proton polarimetry in the GeV region made possible to apply the polarization method suggested by A.I. Akhiezer and M.P. Rekalo at the end of the sixties \cite{Akhiezer:1968ek,Akhiezer:1974em}.
These authors pointed out that the polarization transferred from a longitudinally polarized electron beam to a polarized proton target (or the measurement of the polarization of the recoil proton) in elastic electron proton scattering contains an interference term between the electric and magnetic amplitudes that is more sensitive to a small electric contribution and also to its sign.

Earlier, the privileged method to extract FFs was based to the 'Rosenbluth separation'  \cite{Rosenbluth:1950yq}: the measurement of the unpolarized cross section for a fixed $Q^2$ at different angles. It turns out that this method is limited by the precision on the extraction of the electric FF, at large $Q^2$, as the magnetic contribution is enhanced by a factor of $\tau=Q^2/4M^2$, $M$ being the proton mass.

The data on the FFs ratio, collected mostly by the GEp collaboration at JLab (\cite{Puckett:2017flj} and References therein) show that not only the precision is larger as expected but also that the ratio deviates from unity, as previously commonly accepted. Meaningful data were collected up to
$Q^2\simeq$ 9 GeV$^2$.  A $Q^2$- increasing discrepancy appeared between polarized and unpolarized elastic scattering experiments, giving rise to a large number of publications and speculations. Several issues were discussed: radiative corrections \cite{Bystritskiy:2007hw,Gramolin:2016hjt,Gerasimov:2015aoa}, parameter correlations \cite{TomasiGustafsson:2006pa}, relative normalization within a set of data and among sets of data \cite{Pacetti:2016tqi,Arrington:2003df}, as well as the validity of the one-photon exchange approximation. This last point is of main interest for the present work. It is intended that the two-photon exchange (\TPE) contribution is of the order of $\alpha^2$ ($\alpha=1/137$ is the fine structure constant of the electromagnetic interaction, and $Z$ the target charge number), but the \TPE  contribution  discussed here corresponds to the interference term between one and two photon exchange. Such interference is, in principle, of the order of $\alpha$ \cite{DeRujula:1972te,DeRujula:1973pr} and contains several contributions, as discussed below in more details. Among them, the enhancement  of model dependent terms not included in standard radiative corrections (the 'hard box' contributions, where both virtual photons carry large part of the transferred momentum) has been object of several recent model calculations that are however controversial, and quantitatively disagree at few percent level (for a recent review, see \cite{Afanasev:2017gsk} and References therein).

It is fair to remind that in the 70's the presence of a possible \TPE contribution was under scrutiny of experimental and theoretical investigations \cite{Gunion:1972bj,Boitsov:1972if,Franco:1973uq}. It was theoretically predicted that a possible large effect could arise from \TPE when $Q^2$ increases due to the fact that a reaction mechanism where the transferred momentum is equally shared between the two photons can  compensate the scaling in $\alpha$ due to the steep decreasing of the form factors with  $Q^2$. As a conclusion of a series of measurements (for a review, see \cite{TomasiGustafsson:2009pw}), no experimental evidence was found, in limits of the precision of the data, and, since that time, the one photon exchange approximation was assumed {\it a priori}. Two ($n$)- photon exchange can therefore contribute, although the size of the amplitude is scaled by the factor $Z\alpha$ $((Z\alpha)^n)$. In this context, it is expected that \TPE become more important
\begin{enumerate}
\item when $Q^2$ increases;
\item when the charge $Z$ of the target increases.
\end{enumerate}

Model independent statements, derived from symmetry properties of the strong and electromagnetic interactions, give reliable predictions of the  \TPE contribution to the observables \cite{Rekalo:2003xa,Rekalo:2003km,Rekalo:2004wa}:
\begin{itemize}
\item FFs acquire an imaginary part, and one additional charge-odd amplitude, of the order or $\alpha$, enters in the expression of the current.
\item instead that two FFs , functions only of $Q^2$, these three new amplitudes are complex  functions of two variables $(E, \theta)$ or ($Q^2, \epsilon$), where $ \epsilon=[1+2(1+\tau)\tan^2(\theta/2)]^{-1}$ is the linear polarization of the virtual photon, and $E$ ($\theta$) is the energy (angle) of the scattered electron in the laboratory (lab) system.
\item  non linearities arise in the Rosenbluth fit, i.e., in the unpolarized (reduced) cross section versus $\epsilon$ at fixed $Q^2$.

\item due to the charge-odd (C-odd) terms,  a non vanishing charge--asymmetry should be observed in $e^{\pm} p$  scattering:
\be
A^{odd}=
\displaystyle\frac{\sigma (e^+ p\to e^+p )-\sigma (e^- p\to e^-p )}
{\sigma (e^+ p\to e^+p )+\sigma (e^- p\to e^-p )}.
\label{eq:eqasym}
\ee
\end{itemize}

Summarizing, it can be stated from these general features that, in presence of a sizable \TPE contribution, one expects: - $\epsilon$ non-linearities in the Rosenbluth plot  - a charge asymmetry (differences in $e^\pm p$ elastic cross sections, in the same kinematical conditions) - and non vanishing parity-odd polarization observables. All these effects would increase with $Q^2$.

Note that for the crossed channels (the annihilation channels $e^++e ^-\leftrightarrow p+\bar p$) \TPE effects would be  seen as an asymmetry in the unpolarized angular distribution \cite{Rekalo:1999mt}, $i.e.,$
the presence of odd terms with respect to $\cos\tilde{\theta}$ (where $\tilde{\theta}$ is the center of mass (cms) angle of the produced particle).

In Ref. \cite{Kuraev:2006ys}, an exact QED calculation was performed for $e^{\pm}
\mu ^-$ scattering, and  for the crossed process. This calculation was then applied to $ep$ scattering in Ref. \cite{Kuraev:2007dn}. The obtained charge asymmetry is expressed as the sum of the contribution of two virtual photon exchange, (more exactly the interference between the Born amplitude and the box-type amplitude) and a term from soft photon emission. %In Ref. \cite{Kuraev:2007dn} one should have added a change of sign in the odd terms (from negative muon to positive proton).

In the  total contribution from hard \TPE, in addition to  the contribution of  elastic
proton form factors, intermediate excited proton states should be taken into account \cite{Kuraev:2006ys}.  Based
on  sum rules developed in QED, it is possible to show that these two
contributions are mutually cancelled, and that only the point-like \TPE  should be taken into account for the hard \TPE contribution.
This is also in agreement with some model calculations that find corrections with opposite signs for elastic nucleon and $\Delta$ or $N^*(1535)$ excitation \cite{Tomalak:2017shs}. 

In this work we compile and discuss the results of three recent experiments, that were especially built to detect a possible charge asymmetry  through the measurement of the cross section ratio of electron and positron elastic scattering on the proton. This observable is sensitive to the real part of the \TPE amplitude. The recent data are compared with a calculation \cite{Kuraev:2007dn}, where no specific model dependent  enhancement of the \TPE contribution is added.

%%%%%%%%%%%%%%%%%%%%%%%%%%%%%%%%%%%%%%%%%%%%%%%%%%%%%%
\section{General considerations}
%%%%%%%%%%%%%%%%%%%%%%%%%%%%%%%%%%%%%%%%%%%%%%%%%%%%%%

Assuming one photon exchange, the unpolarized elastic cross section $d\sigma_{el}$ for lepton-hadron elastic scattering in the Born approximation can be expressed  in terms of two structure functions, $A$ and $B$, which depend  on the momentum squared of the transferred photon, $Q^2$, only:
\be
d\sigma_{el} (e^{\pm} h\to e^{\pm}h )=
d\sigma _{Mott}
\left [A(Q^2)+B(Q^2) \tan ^2\frac{\theta}{2}\right ],
\label{eq:eqs}
\ee
where $d\sigma_{Mott}$ is the cross section for point-like particles. This is a very general expressions that holds for any hadron of any spin $S$. The structure functions depend on the $2S+1$ electromagnetic form factors, where $S$ is the spin of the hadron. In the Born approximation, the elastic cross section is identical for positrons and electrons. Two kinematical variables characterize this process, usually the polarization of the virtual photon $\epsilon$ and the momentum transfer squared, $Q^2$  or  the incident energy $E$ and the electron scattering angle $\theta$.

\begin{figure}
\mbox{\epsfxsize=17.cm\leavevmode \epsffile{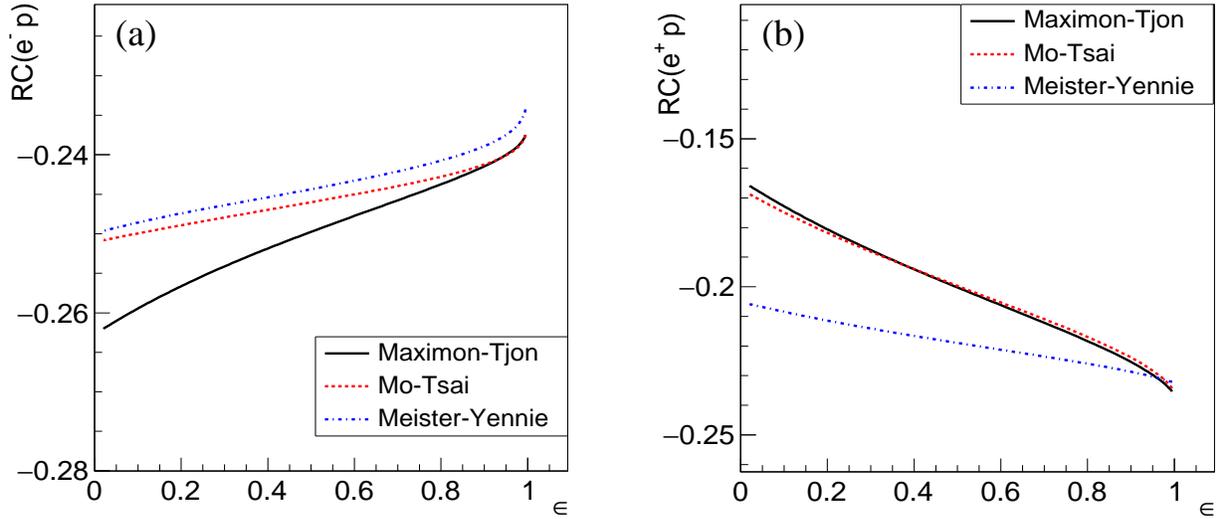}}
%\mbox{\epsfxsize=8.cm\leavevmode \epsffile{RCe-.eps}}
%\mbox{\epsfxsize=8.cm\leavevmode \epsffile{RCe+.eps}}
\caption{Radiative correction factor as a function of $\epsilon$ for $e^-p$  (a)  and $e^+p$ (b), from Ref. \cite{Maximon:2000hm} (solid black line), Ref.  \cite{Mo:1968cg} (dashed red line), Ref. \cite{Meister:1963zz} (dash-dotted blue line) for $Q^2$=1 GeV$^2$ and
$\Delta E =0.01\ E' $.}
\label{Fig:RC}
\end{figure}

Note that the Born elastic cross section is intended to be the measured cross section, $d\sigma_{meas}$ after applying radiative corrections  that take into account photon radiation from the charged particles, $\delta^\pm$. More precisely:
\be
d\sigma_{meas}^\pm=d\sigma_{el}(1+\delta^\pm),
\ d\sigma_{el}= \frac{d\sigma_{meas}^\pm }{(1+\delta^\pm)},
\label{eq:eqB}
\ee
where $\delta^\pm$, besides  charge even terms, contains charge-odd terms (that change sign for positron scattering). The sign $+$ ($-$) stands for scattering on positrons (electrons)):
$\delta^\pm=\mp \delta_{odd}+  \delta_{even}$
%= \delta_0 \mp\delta_1+\delta_2$.
%The indexes $i,$ $i= 0,1, 2$ stand for the terms proportional to $Z^ i$, where $Z$ is the target atomic number.
One can write the odd term $\delta_{odd}$ as the sum of a "hard" ($2\gamma$) and a "soft" ($s$) contributions:
\be
\delta_{odd}=\delta_{2\gamma}+\delta_s.
\label{eq:delta}
\ee
In the experimental works considered here,  only $\delta_s$ was included in the radiative corrections, although the splitting (\ref{eq:delta}) may  differ in different formalisms. Different  calculations were applied to the data considered here, see Refs. \cite{Maximon:2000hm,Mo:1968cg,Ent:2001hm,Gramolin:2014pva}. As an example we illustrate the difference of $\delta^\pm$ from some first order calculations in Fig. \ref{Fig:RC}a for electron  and in Fig. \ref{Fig:RC}b for positron scattering, as a function of $\epsilon$ for $Q^2$=1 GeV$^2$. The soft corrections depend from the inelasticity parameter $\Delta E$ taken here as 1\% of the scattered energy, $E'$. The difference among the calculations is of the order of few percent, depending on $\epsilon$,  $Q^2$ and $\Delta E$. Note that a larger value, $\Delta E\simeq 0.03 E'$, is closer to the typical experimental cut, but a smaller value enhances the effect and is taken here for illustration.
This suggests that effects of the order of percent may be attributed to different procedures of  applying odd soft corrections to the data.

A deviation from unity of the ratio:
\be
R^{meas}=
\displaystyle\frac{d\sigma^{meas} (e^+ p\to e^+p )}
{d\sigma^{meas} (e^- p\to e^-p )} =\frac{1+\delta_{even}-\delta_{2\gamma}-\delta_s}{1+\delta_{even}+\delta_{2\gamma}+\delta_s} 
\label{eq:ratio}
\ee
is a clear signature of (soft and hard) charge-odd contributions to the cross section.

A C-odd effect is enhanced in the ratio of $e^+ p\to e^+p$ over $e^- p\to e^-p$ cross sections, $R$,  with respect to the asymmetry, $A^{odd}$:
\be
A^{odd}=
\displaystyle\frac{d\sigma (e^+ p\to e^+p )-d\sigma (e^- p\to e^-p )}
{d\sigma (e^+ p\to e^+p )+d\sigma (e^- p\to e^-p )}=\frac{\delta_{odd}}{1+\delta_{even}}=\frac{R-1}{R+1},\
R= \frac{1+A_{odd}}{1-A_{odd}}.
\label{eq:eqasym1}
\ee

%This sign was corrected
In  Eq. (5) of Ref. \cite{TomasiGustafsson:2009pw}, a (small) correction was added in the asymmetry, taking into account the even radiative corrections. This correction is indeed small, but depends on the elasticity cut and on the way the radiative corrections were implemented.  $\delta_{even} $ from Ref.  \cite{Maximon:2000hm} was implemented. The results were given for an inelasticity cut $\Delta E/E = 0.03$, that is consistent with most experiments. Let us stress however that our result for the hard $2\gamma$ contribution does not depend on this term and on the cut.

The charge asymmetry that includes soft and hard $2\gamma$  contributions at first order in $\alpha$ is calculated in Ref. \cite{Kuraev:2007dn}:
\ba
A_{odd}^K&=&\frac{d\sigma^{e+p}-d\sigma^{e^-p}}{d\sigma^{e+p}+d\sigma^{e^-p}}
 =\frac{2\alpha}{\pi(1+\delta_{even})}
\biggl[ \ln \frac{1}{\rho}
\ln\frac{(2\Delta E)^2}{ME}-\frac{5}{2}\ln^2\rho +
\ln x\ln\rho+\nn \\
&&
\Li\left (1-\frac{1}{\rho x}\right)-\Li\left(1-\frac{\rho}{x}\right )\biggr ],
\label{eq:eqEAK}
\ea
$$\rho =\left (1-\frac {Q^2}{s}\right)^{-1}=1+2\frac{E}{M}\sin^2\frac{\theta}{2},
~x=\frac{\sqrt{1+\tau}+\sqrt{\tau}}{\sqrt{1+\tau}-\sqrt{\tau}}.$$
The term containing $\Delta E$ gives the largest contribution to the asymmetry and has a large $\epsilon$ dependence.

By  correcting the data for  the contributions of the vertex-type corrections $\delta_{even}$ and soft two-photon contributions $\delta_s$, $R^{meas}$  from Eq. (\ref{eq:ratio}) reduces to
\be
R_{2\gamma}\simeq \frac{1-\delta_{2\gamma}}{1+\delta_{2\gamma}},
\label{eq:eq2gamma}
\ee
where $\delta_{2\gamma}$ is the contribution of hard virtual two-photon exchange. Building the ratio $R_{2\gamma}$ enhances those contributions to the two-photon amplitudes that depend on off-mass shell proton states.

The data on $R_ {2\gamma}$ have been corrected for those radiative corrections  that depend on the inelasticity cut and contain the term proportional to $\Delta E$. The largest odd contribution, indeed, arises from this term. In order to compare the results from different experiments, it would be wise to use the same ansatz for radiative corrections, what turns out not to have been the case. Therefore, we must take into account that a difference of 1 or 2\% in the data may be attributed to the different corrections. The issue of the approximations used in the past, where mainly first order radiative corrections were considered \cite{Meister:1963zz,Mo:1968cg,Tsai:1961zz,Maximon:2000hm}, has been recently discussed in a series of articles \cite{Gramolin:2016hjt,Gerasimov:2015aoa,TomasiGustafsson:2009pw} as well as in a recent review \cite{Pacetti:2015iqa}, whereas the role of higher order corrections was pointed out in \cite{TomasiGustafsson:2006pa,Bystritskiy:2007hw}.

The odd radiative correction term is usually  splitted in the following parts:
\begin{enumerate}
\item Bremstrahlung process, with emission of a  real photon: this part of contribution is strictly depend on the experimental cuts over measured energy and angles of detected particles. This term is large, and contains the infrared singularities which cancels with one from virtual two photon contribution.
\item virtual two photon corrections, which are splitted in two parts due to the uncertainties of the calculation with respect to proton form factors and intermediate proton state contributions:
\begin{enumerate}
\item soft part of two photon virtual contribution, that includes the case when one of the virtual photon in soft. In this case the intermediate proton and electron are almost on mass shell,  and one can treat this term as one photon exchange contribution, with some factor of additional soft virtual photon. This part of contribution can be exactly calculated in  QED, and contains infrared singularities which cancels with the real soft photon contribution;
\item hard part, where both virtual photon are hard. In this case one has to consider six proton form factors instead of two, where one of the protons is off-shell, and in addition,  some intermediate proton states, as $\Delta$ resonance, etc. This part of contribution  is strictly dependent  over different theoretical assumptions  and is the object of the experimental measurements.
\end{enumerate}
\end{enumerate}	
The splitting of the two photon contribution into the soft and hard parts is not uniquely defined and may differ from one author  to another. The answers differ by some finite expression, which depends on kinematical invariants, and can be explained by different methods of calculation.  The generally adopted  approach is the splitting that was considered in the works of two groups \cite{Maximon:2000hm,Mo:1968cg}. The soft part of  two-gamma contribution, calculated by Mo Tsai generally used in the experiments is \cite{Mo:1968cg}:
\begin{eqnarray}
&\delta_{soft Tsai}&=-\frac{\alpha}{\pi} \left [2\ln\rho  \left (2\ln\frac{E}{\Delta E}-3 \ln\rho\right ) \right ]
+Li_2  \left  (-\frac{M-E'}{E} \right  )-Li_2 \left  [\frac{M(M-E')}{E_4 E' -M E}\right  ]
\label{eqMTs} \\
&&+Li_2  \left [\frac{2E'(M-E')}{2 E_4E'-ME}\right ]+\ln  \left  |\frac{2E_4 E'-M E}{E(M-2E')} \right |\ln\frac{M}{2E'}-Li_2 \left (-\frac{E_4-E'}{E'} \right )
\nonumber \\
&&+Li_2 \left [\frac{M(E_4-E')}{2E_4 E -M E'}\right ]-Li_2 \left [\frac{2E(E_4-E')}{2 E_4E-ME'} \right ]-\ln  \left  |\frac{2E_4 E-M E'}{E'(M-2E)} \right |\ln\frac{M}{2E'}
\nonumber \\
&&-Li_2 \left  (-\frac{M-E}{E} \right )+Li_2 \left  (\frac{M-E}{E} \right )-Li_2 \left  [\frac{2(M-E)}{M} \right ]-\ln  \left |\frac{M }{2E'-M} \right |\ln\frac{M}{2E}
\nonumber \\
&&+Li_2 \left  (-\frac{M-E'}{E'}\right )-Li_2 \left  (\frac{M-E'}{E'}\right )+Li_2  \left  [\frac{2(M-E')}{M}\right ]+\ln  \left |\frac{M }{2E'-M}\right |\ln\frac{M}{2E'}.
\nonumber 
\end{eqnarray}
Here  $E'(E_4)$ is the energy of scattered electron(proton) $ E_4=E+M-E'$, and $M$ is the proton mass, $Li_2$ is the Spence function.
%%%%%%%%%%%%%%%%%%%%%%%%%%%%%%%%%%%%%%%%%%%%%%%%%%%%%%%%%%%%%%%%%%%%%%%
A similar but not equal expression is given in Eq. (5.2) for the work of Ref. \cite{Maximon:2000hm}.

%%%%%%%%%%%%%%%%%%%%%%%%%%%%%%%%%%%%%%%%%%%%%%%%%%%%%%
\section{Compilation of  recent $e^{\pm} p$ results}
%%%%%%%%%%%%%%%%%%%%%%%%%%%%%%%%%%%%%%%%%%%%%%%%%%%%%%

\begin{figure}
  \begin{center}
    \includegraphics[width=0.98\columnwidth]{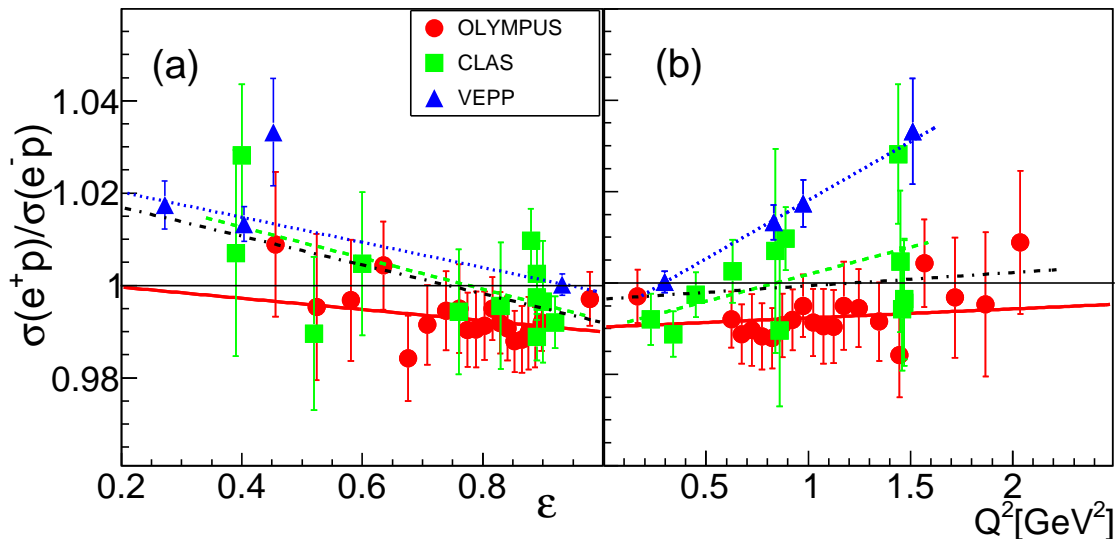}
  \caption{Radiatively corrected ratio of positron to electron cross sections $R_ {2\gamma}=\sigma(e^+p)/ \sigma(e^-p)$ with the corresponding linear fits
 as a function of $\epsilon$ (a) and $Q^2$ (b) from OLYMPUS \cite{Henderson:2016dea} (red circles and red solid line), CLAS \cite{Rimal:2016toz} (green squares and green dashed line)  and VEPP-3 \cite{Rachek:2014fam} (blue triangles and blue dotted line). The black dash-dotted line corresponds to the global linear fit.}
 \label{Fig:Alldata}
    \end{center}
\end{figure}

Three results from recent experiments that measured the ratio $R_ {2\gamma}$ from a radiatively corrected cross section ratio, $R^{meas}$, Eq. (\ref{eq:ratio}),  are shown  as a function of  $\epsilon$in Fig. \ref{Fig:Alldata}(a)  and  as a function of $Q^2$ in Fig. \ref{Fig:Alldata}(b), with the corresponding linear fits. Most of the data deviate from unity by less than 2\%. A slight increase with decreasing $\epsilon$ is seen. The authors of the VEPP experiment, Ref. \cite{Rachek:2014fam},  point out a significant \TPE effect increasing with $Q^2$,  but this is not confirmed by the OLYMPUS  data \cite{Henderson:2016dea}. For OLYMPUS, when not explicitly mentioned, the data set (a) i.e, corrected from Mo-Tsai to order $\alpha^3$ are considered. Note that, however, the VEPP results, that are the most precise, lack an absolute normalization. 
The weighted average of the ratio $R_ {2\gamma}$ for all data and for the individual data set, to be compared to unity for no \TPE  contribution, is shown in Table \ref{Table:Table1}. The  compatibility with a constant $R_ {2\gamma}=1$ is  indicated  by  $\chi ^2/N(1)$.  One may see that a deviation of about $3\sigma$ is visible for the VEPP data, where $\chi^2/N$ is much larger than 1. The average for this data set is larger than unity, whereas it is smaller for the CLAS and OLYMPUS data. Adding a parameter decreases the $\chi ^2/N$, that falls below unity for these two last sets of data. The results of the linear fit are also reported in Table \ref{Table:Table1}. The fact that $\chi ^2/N\simeq 2.0$ for all data sets is much larger than for each individual set shows the large difference between the VEPP data compared to the two others data sets.
\begin{table}
\begin{tabular}{|c|c|c|c|c|c|}
\hline\hline
\multicolumn{2}{|c|}{ }   & All data & OLYMPUS & CLAS  & VEPP   \\
\hline
   \multicolumn{6}{|c|}{Experiment}\\
\noalign{\smallskip}\hline
%\multicolumn{6}{|c|}{Experiment}\\
\multicolumn{2}{|c|}{$<R_{2\gamma}>$    }       & 0.999$\pm$  0.001 &  0.999$\pm$ 0.001 & 0.997 $\pm$  0.002 & 1.006$\pm$ 0.002 \\
 \multicolumn{2}{|c|}{      $\chi ^2/N$(1) }      &  69.3/35=1.98         &19/19=1.00       & 12.1/11=1.1         & 23.7/3=7.9 \\
    \hline
\multirow{2}{*}{$R_{2\gamma}=a_0+a_1 \epsilon$}    
  &$a_0$      &1.023$\pm$  0.005   & 1.002$\pm$  0.014 &  1.026$\pm$ 0.018 & 1.026 $\pm$  0.005  \\
  &$a_1$    & -0.031$\pm$   0.006& -0.012$\pm$  0.017 &  -0.034$\pm$ 0.020 & -0.027 $\pm$  0.007  \\
 % \multicolumn{2}{|c|}{    $\chi ^2/N$    } &38.6/34= 1.13& 5.44/18= 0.3   &9.72/10 =0.97     & 3.08/2 =1.5            \\
 \cline{2-6}
\multicolumn{2}{|c|}{  $\chi ^2/N$ }    &38.6/34= 1.13& 5.44/18= 0.3   &9.72/10 =0.97     & 3.08/2 =1.5            \\
    \hline
 \multirow{2}{*}{$R_{2\gamma}=b_0+ b_1 Q^2$ }       
    &$b_0$& 0.981$\pm $0.004 &0.990$\pm$  0.005 &  0.990$\pm$ 0.004 & 0.992 $\pm$  0.004  \\  
   &$b_1$&  0.014$\pm$  0.003 &0.002$\pm$  0.005 &  0.011$\pm$ 0.006 & 0.026 $\pm$  0.006  \\
 \cline{2-6}
\multicolumn{2}{|c|}{ $\chi ^2/N$ }     &68.3/34=  2.0   &5.74/18=0.32       & 8.4/10=0.8   & 0.05/2=0.02 \\
    \hline
\end{tabular}
\caption{Weighted average of the ratio $R_ {2\gamma}$ for all data and for the individual data sets (the OLYMPUS data corresponding to the set (a) of Ref. \cite{Henderson:2016dea}), to be compared to unity for no $2\gamma E $ contribution. The   compatibility with a constant $R_ {2\gamma}=1$ is  indicated  by  $\chi ^2$(1). The results from linear fits in $\epsilon$ and $Q^2$ are also given.}
\label{Table:Table1}       % Give a unique label
\end{table}

In the analysis of the experimental data, the radiative correction codes are embedded in the MonteCarlo used to analyze the data, and it is not straightforward to unfold the effects from the acceptance and the efficiency of the setup.  Note that the $\Delta E$ term is by far the most sizable among the odd terms, becoming larger when the inelasticity cut is smaller. At the elastic peak it becomes infinite.

It may be  difficult to evaluate the size of the applied radiative corrections and their dependence on the relevant kinematical variables from the published results. However, it is always possible to calculate radiative corrections for different energy cuts to compare different models and study, at least qualitatively, their effect in comparison to the data.  In particular we consider the calculations from  Refs. \cite{Maximon:2000hm} and \cite{Mo:1968cg}, that were most often used in the experimental papers.

As radiative corrections applied to the data may differ from one paper to another  by some finite expression (which depends on kinematical invariants), in order to be less sensitive to model corrections, we consider the total odd contribution from Ref. \cite{Kuraev:2007dn} and remove the odd correction from the calculations used in the data. This means that we have to proceed from  $R^{meas}$ to  $R_{2\gamma}$:
\be
R_{2\gamma}^{K}=\frac{1-A^K_{odd}(1+\delta_{even})+\delta_{M}}{1+A^K_{odd}(1+\delta_{even})-\delta_{M}},
\label{eq:ratioc}
\ee
where $\delta_{M} $ can be calculated  from Ref. \cite{Mo:1968cg}, here reported in Eq. (\ref{eqMTs}), or from the corresponding correction of Ref. \cite{Maximon:2000hm}.
%, or from the 2nd line of Eq. (5.2) of Ref. \cite{Maximon:2000hm}.

We calculate the asymmetry $A^K_{odd}$ from Eq. (\ref{eq:eqEAK}), then the ratio $R_{2\gamma}^{K}$ from Eq. (\ref{eq:ratioc}) to be compared to the data. The ratio depends on two variables, $Q^2$ and $\epsilon$.  First we study the $Q^2$ and $\epsilon$ dependence separately, then, in order to have all the data and the calculation in a plot, we consider the absolute difference between each data point and Eq. (\ref{eq:eqEAK}), calculated for the corresponding values of the two variables, $Q^2$ and $\epsilon$.

%%%%%%%%%%%%%%%%%%%%%%%%
\subsection{The VEPP experiment}
%%%%%%%%%%%%%%%%%%%%%%%%

The experiment, published in Ref. \cite{Rachek:2014fam,Nikolenko:2015xsa}, was performed at the VEPP-3 storage ring. The
$e^\pm p$ cross section was measured for two beam energies, 1.6 and 1 GeV and different lepton scattering angles, spanning such $\epsilon,Q^2$ kinematical ranges:   $0.272 < \epsilon < 0.932$, and $0.298< Q^2 <1.0332 $ GeV/c$^2$.

\begin{figure}
\mbox{\epsfxsize=18.cm\leavevmode \epsffile{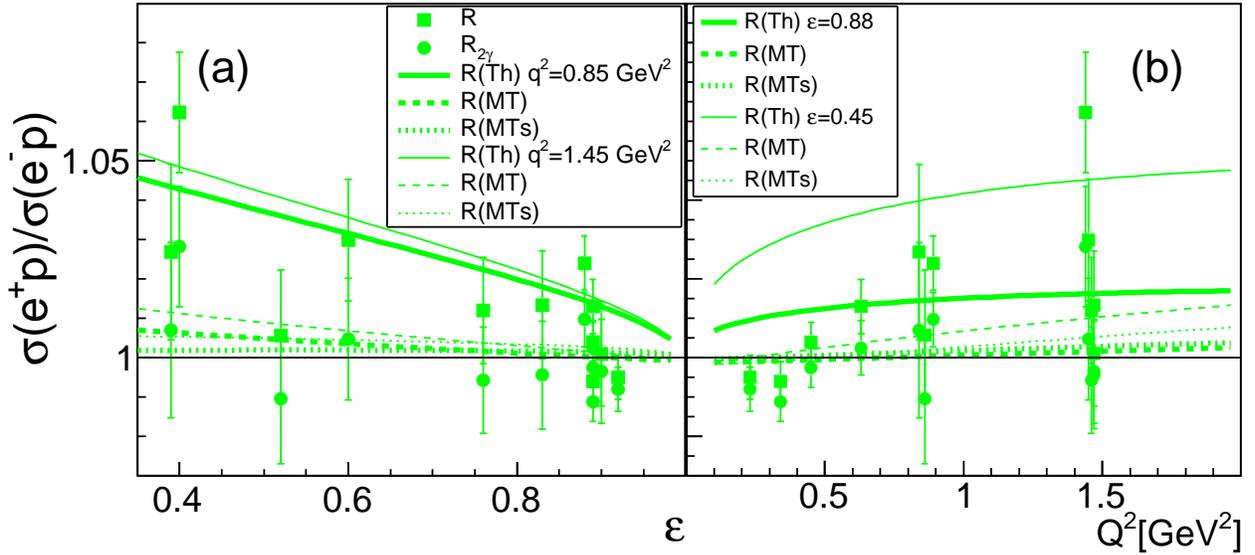}}
\caption{Ratio of cross sections $\sigma(e^+p)/ \sigma(e^-p)$ as a function of $\epsilon$ (a) and of $Q^2$ (b) from VEPP-3
\protect\cite{Rachek:2014fam} : raw ratio, $R$, (blue  squares) and  soft corrected ratio $R_{2\gamma}$ (blue circles). Solid(open) symbols correspond to $E\simeq$1.6(1) GeV. The calculation from  Ref.
\protect\cite{Kuraev:2007dn}, Eq. (\ref{eq:eqEAK}) is shown as a solid line, after correction for the soft contribution from  \protect\cite{Maximon:2000hm} (dashed line)
and  from Ref. \protect\cite{Mo:1968cg} (dotted line)  (see text). Thick(thin) lines correspond to E=1.6(1) GeV. }
\label{Fig:Vepp}
\end{figure}

The measured (uncorrected) ratio $R$  is shown  as solid blue squares in Fig.  \ref{Fig:Vepp} for the VEPP-3 experiment \cite{Rachek:2014fam}. They correspond to the raw ratio, before applying radiative corrections. The deviation from unity is due to the odd terms, that are due to soft corrections  and hard \TPE terms. The ratio of the radiatively corrected cross section, $R_{2\gamma}$ (blue circles), deviates from unity only in the presence of a hard two photon contribution not included in the radiative corrections. The radiative corrections applied to the data are based on the ESEPP generator developed in Ref. \cite{Gramolin:2014pva} (dot-dashed line).

In order to compare separately the $\epsilon$ and $Q^2$ dependencies, we report in Fig. \ref{Fig:Vepp} the calculation from Ref. \cite{Kuraev:2007dn} fixing $Q^2$ to an average value of 1 GeV$^2$ (a) and $\epsilon=0.4$ in the right side (b). The total odd contribution is in good agreement with the uncorrected experimental ratio. The hard $2\gamma$ contribution is shown after subtraction from Ref. \cite{Kuraev:2007dn} of the soft correction from Ref. \cite{Mo:1968cg} (dotted line) and  from  \cite{Maximon:2000hm} (dashed line). The quantitative effect is discussed below.

%%%%%%%%%%%%%%%%%%%%%%%%
\subsection{The OLYMPUS experiment}
\begin{figure}
\mbox{\epsfxsize=18.cm\leavevmode \epsffile{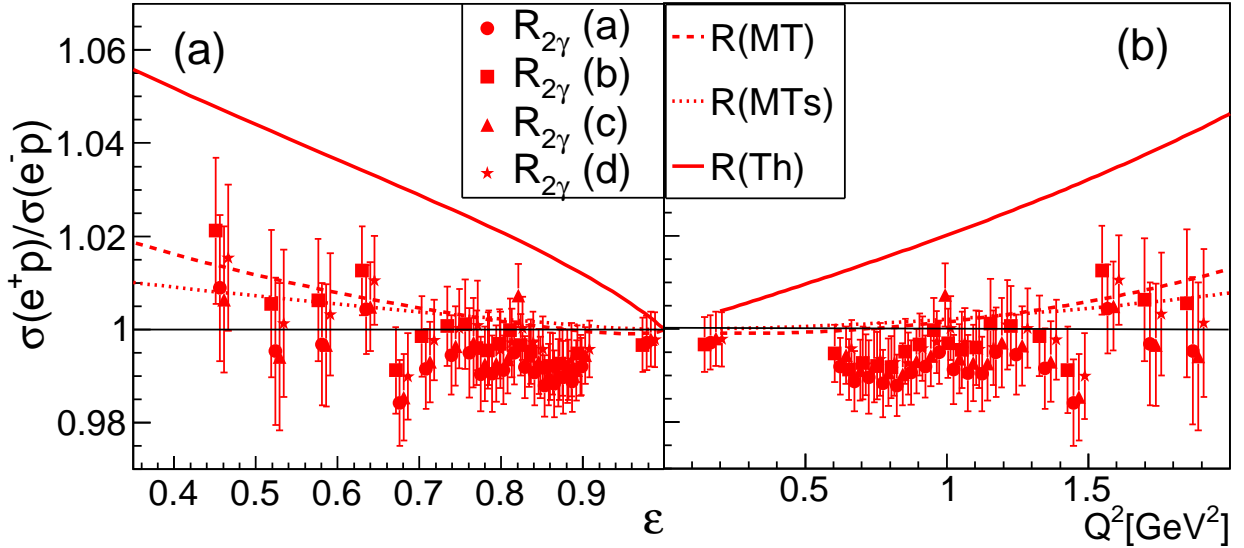}}
%\vspace*{.2 truecm}
\caption{Ratio of cross sections $R=\sigma(e^+p)/ \sigma(e^-p)$, as a function of $\epsilon$ (a) and of $Q^2$ (b) and the calculation from Eq. (\ref{eq:eqEAK}) for $E$=2.01 GeV. The considered experiment is  OLYMPUS \cite{Henderson:2016dea} (red circles, squares, triangles and stars). Lines as in Fig.  \protect\ref{Fig:Vepp}.}
\label{Fig:Olympus}
\end{figure}

%%%%%%%%%%%%%%%%%%%%%%%%
This experiment, published in Ref. \cite{Henderson:2016dea}, was performed at the DORIS storage ring at DESY, using 2.01 GeV electron and positron beams impinging on an internal hydrogen gas target.  Twenty values of the ratio $R$ were measured in the range:  $0.456 < \epsilon < 0.978$, and $0.165< Q^2 <2.038$ GeV/c$^2$.
Most of these values lie within $|R|<1.02 $ with a mild tendency to increase at large $Q^2$ and/or small $\epsilon$. The $\epsilon$ ($Q^2$)-dependence  of the data is shown in  Fig. \ref{Fig:Olympus}a (Fig. \ref{Fig:Olympus}b).

Four options of radiative corrections were implemented, following Mo-Tsai at first order (solid red circles) (a) or including approximately high orders by exponentiation (solid red squares) (b), and following Maximon-Tjon \cite{Maximon:2000hm} at first order (solid red triangles) (c) or exponentiation (solid red stars) (d). The difference among these options is a few per-thousand and induces at most a difference of 1.5 \% in the extracted ratios.

The statistical error is evaluated to be $<$ 1 \% and the systematical error is $<$ 1.5 \%, the largest source being attributed to the selection of the elastic events. As we are interested in a global difference, for the comparison with the calculation we report the points of Table II of Ref. \cite{Henderson:2016dea}, with systematic, statistical correlated and uncorrelated errors summed in quadrature.  

The calculation is shown in Fig. \ref{Fig:Olympus} for the corresponding fixed beam energy, E=2.01 GeV. 

The  total odd contribution from Ref. \cite{Kuraev:2007dn}  is shown as a red solid line, and after subtraction of the soft correction from Ref. \cite{Mo:1968cg} (dotted line) and  from  \cite{Maximon:2000hm} (dashed line).

The calculations after subtraction, for both choices of the radiative correction ansatz, fall within the errors of most data points. 

%%%%%%%%%%%%%%%%%%%%%%%%
\subsection{The CLAS experiment}
%%%%%%%%%%%%%%%%%%%%%%%%

 The CLAS experiment \cite{Rimal:2016toz} published a list of 19 points of the ratio, for two $Q^2$ values, 0.85 and 1.45 GeV$^2$ and several $\epsilon $ values in the range $0.39 < \epsilon < 0.91 $. The electron and positron where produced by converting a photon beam into $e^\pm$ pairs, which explains partly the largest uncertainty of these data compared to the two previous experiments. Overlapping kinematics reduce the set to 12 independent data points, for comparison to the other data and calculations. The data were radiatively corrected following Ref. \cite{Ent:2001hm}. It is a first order calculation, developed for inelastic scattering to be implemented in MonteCarlo programs and it is based on similar approximation as Ref.  \cite{Mo:1968cg}.

\begin{figure}
\mbox{\epsfxsize=18.cm\leavevmode \epsffile{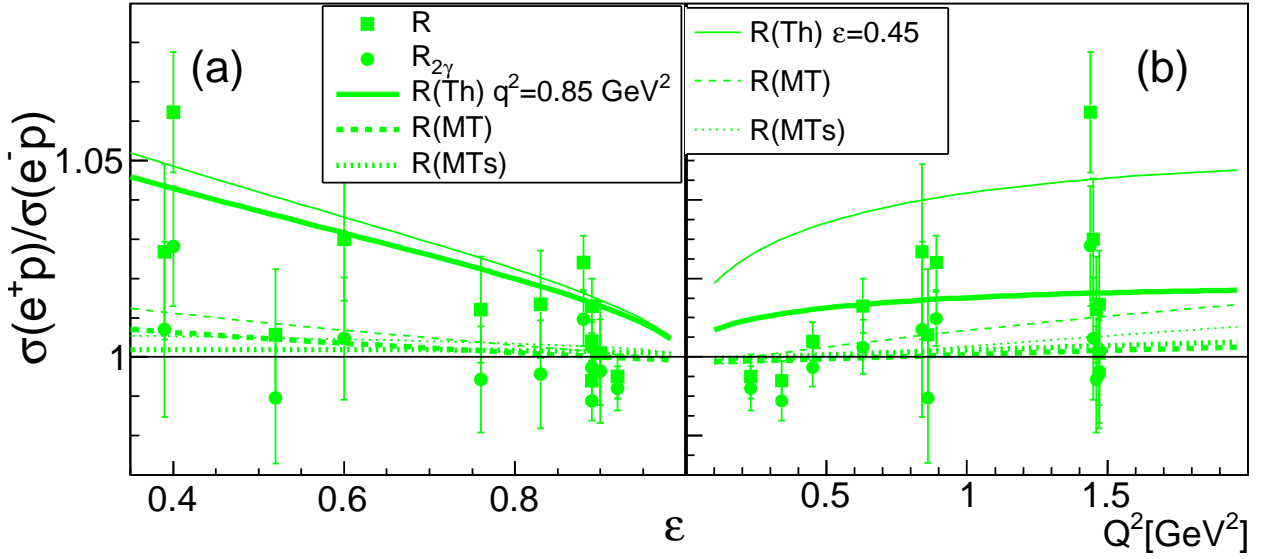}}
%\vspace*{.2 truecm}
\caption{Ratio of cross sections $R=\sigma(e^+p)/ \sigma(e^-p)$, as a function of $\epsilon$ (a) for two values of $Q^2$: $Q^2=0.85$ GeV$^2$ (thick lines)  $Q^2= 1.45$ GeV$^2$ (thin lines)  and as a function of  $Q^2$ (b) for $\epsilon$=0.88 (thick lines) and   $\epsilon$=0.45 (thin lines). The considered experiment is   CLAS \cite{Rimal:2016toz} (green circles and squares). Lines as in Fig.  \protect\ref{Fig:Vepp}. }
\label{Fig:CLAS}
\end{figure}

The data are plotted in Fig. \ref{Fig:CLAS} for the ratio $R$ (solid green squares) and the corresponding calculation from Ref. \cite{Kuraev:2007dn} as a solid green line. For the corrected ratio $R_{2\gamma} $   (solid green circles), the calculation is shown after subtraction of the soft correction from Ref. \cite{Mo:1968cg} (dotted line) and  from  \cite{Maximon:2000hm} (dashed line). The two average values of $Q^2$=0.85 GeV$^2$ and $Q^2$=1.45 GeV$^2$ are considered for the $\epsilon$ dependence, and of  $\epsilon=0.88$ and 0.45  for  the $Q^2$ dependence.

Also in this case the calculations after subtraction, for both choices of the radiative correction ansatz, fall within the errors of most data points.

%%%%%%%%%%%%%%%%%%%%%%%%%%%%%%%%%%%%%%%%%%%%%%%%%%
\section{ Point by point comparison}
\begin{figure}
\mbox{\epsfxsize=18.cm\leavevmode \epsffile{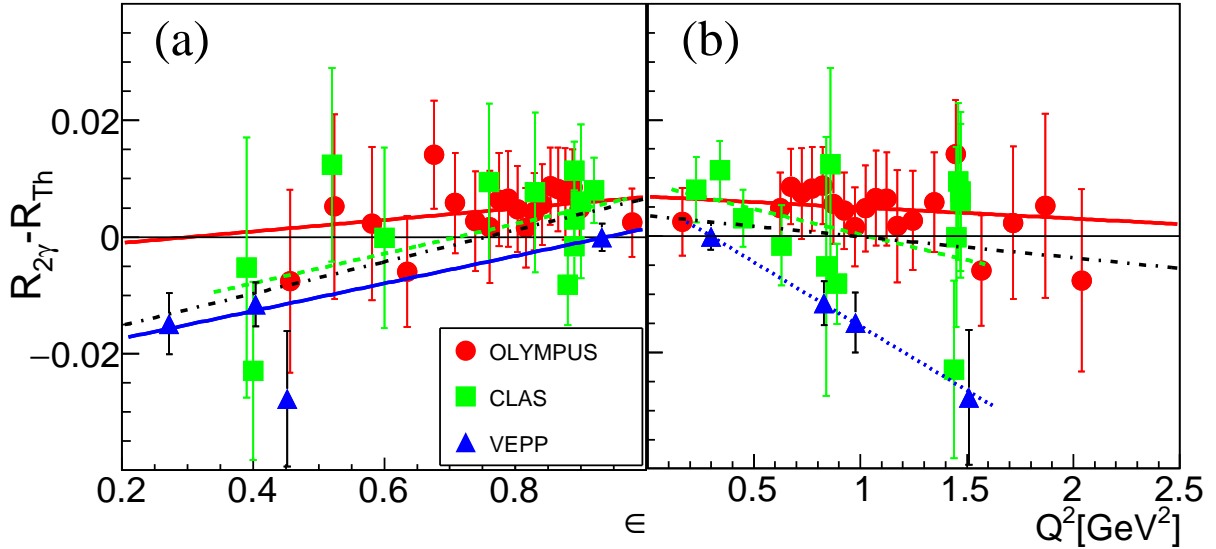}}
\caption{Point to point difference between the calculation from Eq. (\ref{eq:eqEAK}), Ref. \cite{Kuraev:2007dn} and the data for the ratio $R$,  with the corresponding linear fits as a function of $\epsilon$ (a) and $Q^2$ (b) from OLYMPUS \cite{Henderson:2016dea} (red circles and red solid line), CLAS \cite{Rimal:2016toz} (green squares and green dashed line)  and VEPP-3 \cite{Rachek:2014fam} (blue triangles and blue dotted line), after removing the odd corrections as in Eq. (\ref{eq:ratioc}) with $\delta_M$  from Ref. \cite{Maximon:2000hm}. The black dash-dotted line corresponds to the global linear fit.
}
\label{Fig:AllDiff}
\end{figure}

\begin{figure}
\mbox{\epsfxsize=18.cm\leavevmode \epsffile{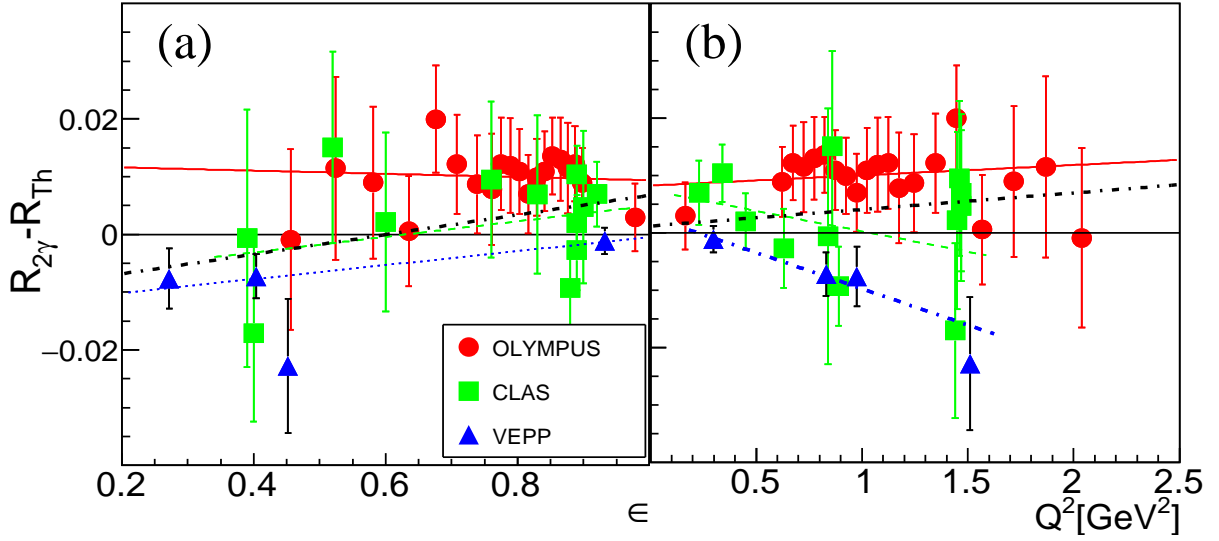}}
\caption{Same as Fig. \ref{Fig:AllDiff}, where $\delta_M$  is calculated from Ref. \cite{Mo:1968cg}. 
}
\label{Fig:AllDiffMTS}
\end{figure}

%%%%%%%%%%%%%%%%%%%%%%%%%%%%%%%%%
The previous analysis wants to evidence the general $\epsilon$ and $Q^2$ dependence of data and calculations,  showing, when available, the full asymmetry as well as the asymmetry after subtraction of radiative corrections.  As the ratio depends on both variables $\epsilon$ and $Q^2$  the difference point by point from the experimental value and the calculation is considered here, which is more rigorous than taking an average value of one variable when plotting along the second one.  The results are shown in Fig. \ref{Fig:AllDiff}a (Fig. \ref{Fig:AllDiff}b for the $\epsilon$($Q^2$) dependence, where the calculation  from Eq. (\ref{eq:eqEAK}), Ref. \cite{Kuraev:2007dn} is plotted  after removing the odd corrections as in Eq. (\ref{eq:ratioc}) with $\delta_M$  from Ref. \cite{Maximon:2000hm}. Similarly for Fig. \ref{Fig:AllDiffMTS}, where the procedure is applied with  $\delta_M$ calculated from Ref. \cite{Mo:1968cg}. 

The calculation of the hard $2\gamma e$ extracted in this way is consistent with the data within the errors. A $Q^2$-dependent discrepancy appears in the data from VEPP, that have a different sign than the other experiments. To quantify the difference between data and calculations we report in Table \ref{Table:Table1a} the average values of this difference for all data as well as for the individual data set and the $\chi ^2$ for a least squares fit. The average ratio is compatible with one, within the error, except for the VEPP data that show also a $\chi^2$ very different for unity. The linear fit finds a mild positive slope for the CLAS and VEPP data, and an intercept compatible with unity at percent level. The $\chi^2\ll 1$ indicates that the number of parameters may be redundant : a two parameter fit may exceed in some cases the precision of the data. 

The difference point by point between data and theory shows in general a discrepancy at per-thousand level in most cases, what is beyond the theoretical and experimental precisions, with a slight $\epsilon$ and $Q^2$ dependencies. 

\begin{table}
\begin{tabular}{|c|c|c|c|c|c|}
\hline\hline
\multicolumn{2}{|c|}{ }   & All data & OLYMPUS & CLAS  & VEPP   \\
\hline
    \multicolumn{6}{|c|}{Difference Theory-Experiment}\\
\noalign{\smallskip}\hline
\multicolumn{2}{|c|}{$<R_{Th}-R_{2\gamma}>_{MTj}$} & 0.003$\pm$ 0.001 &  0.009 $\pm$ 0.002 & 0.003 $\pm$ 0.002 & -0.004$\pm$0.002\\
  \multicolumn{2}{|c|}{  $\chi ^2/N$(0)  }              & 64.1/35=1.83                       &  33.2/19=1.75                       & 10.0/11=0.91     &7.5/3=2.5                                   \\
  \hline
\multirow{2}{*}{$( R_{Th}-R_{2\gamma})_{MTj} =c_0+c_1 \epsilon$} 
&$c_0$& -0.002 $\pm$ 0.005 &-0.003$\pm$ 0.014 &-0.018$\pm$ 0.018   &-0.022$\pm$ 0.005   \\
&$c_1$& 0.027$\pm$ 0.006    &0.009$\pm$ 0.017 &   0.025 $\pm$ 0.021& 0.024 $\pm$ 0.007 \\
 \cline{2-6}
 \multicolumn{2}{|c|}{$\chi ^2/N$}     &    24.1/34= 0.72    & 4.27/18=0.24 &  8.85/10 =0.89 & 2.1/2=1.05                                      \\
  \hline
\multirow{2}{*}{$(R_{Th}-R_{2\gamma})_{MTj} =d_0+d_1 Q^2$} 
&$d_0$ &0.003$\pm$0.002  &0.007$\pm$ 0.005 & 0.009$\pm$ 0.004 &  0.007$\pm$ 0.003 \\
&$d_1$& -0.004$\pm$ 0.003&-0.002$\pm$ 0.005 &-0.009$\pm$ 0.006 &-0.022$\pm$ 0.006  \\
 \cline{2-6}
  \multicolumn{2}{|c|}{$\chi ^2/N$ }    &  46.6/34= 1.37 & 4.4/18= 0.24         &  8.17/10=0.82                       & 0.01/2=0.005                       
                  \\
\noalign{\smallskip}\hline
\noalign{\smallskip}\hline
\multicolumn{2}{|c|}{$<R_{Th}-R_{2\gamma}>_{MTs}$} & 0.002$\pm$ 0.001 &  0.002 $\pm$ 0.001 & 0.004 $\pm$ 0.002 & -0.005$\pm$0.002\\
  \multicolumn{2}{|c|}{         $\chi ^2/N$(0) }    & 59.5/35=1.7  &  23.9/19=1.26     & 11.9/11=1.08   &17.34/3= 5.78           
  \\  \hline
\multirow{2}{*}{$( R_{Th}-R_{2\gamma})_{MTs} =e_0+e_1 \epsilon$} 
&$e_0$&-1.025$\pm$ 0.005 &0.012$\pm$ 0.014 &-0.009$\pm$ 0.018 &-0.013$\pm$ 0.005   \\
&$e_1$&0.017 $\pm$ 0.006 &-0.003$\pm$ 0.017  &   0.014 $\pm$ 0.021& 0.012 $\pm$ 0.007 \\
\cline{2-6}
  \multicolumn{2}{|c|}{ $\chi ^2/N$ }    &  40.2/34=  1.18  & 5.2/018 = 0.29  &  8.77/10=0.88  & 1.95/2=0.9 \\  
\noalign{\smallskip}\hline
\multirow{2}{*}{$(R_{Th}-R_{2\gamma})_{MTs} =f_0+f_1 Q^2$} 
&$f_0$ & 0.001$\pm$0.002&0.008$\pm$ 0.005 & 0.007$\pm$ 0.004 &  0.007$\pm$ 0.004 \\
&$f_1$& 0.003$\pm$0.003&0.002$\pm$ 0.005 &-0.007$\pm$ 0.006 &-0.013$\pm$ 0.006  \\
 \cline{2-6}
 \multicolumn{2}{|c|}{ $\chi ^2/N$}     &  48.5/ 34=1.42& 5.1/18= 0.28     &  7.8/10=0.8  & 0.47/2 =0.2  \\             
\noalign{\smallskip}\hline\hline
\end{tabular}
\caption{Weighted average of the difference  $R_{Th}-R_{ 2\gamma}$ for all data  and for the individual data set (the OLYMPUS data corresponding to the set (a)), to be compared to zero for no $2\gamma e $ contribution. The   compatibility with a constant $R_{Th}-R_{2\gamma}=0$ is given by  $\chi ^2$/N (0). The difference between the data and the calculation of Ref. \cite{Kuraev:2007dn} is given after subtraction of the soft corrections  from  \cite{Maximon:2000hm} and from Ref. \cite{Mo:1968cg}. The results from linear fits in $\epsilon$ and $Q^2$ are also given.}
\label{Table:Table1a}       % Give a unique label
\end{table}

For the OLYMPUS data, the sensitivity to different ansatz of radiative corrections  is shown in Table \ref{Table:Table2}. Also the weighted average of the ratio $R_ {2\gamma}$ for all data slightly changes according to the four different ansatz used to extract the data. The difference between theory and data for the different radiative corrections  is shown in Table \ref{Table:Table2a}. 
\begin{table}
\begin{tabular}{|c|c|c|c|c|c|}
\hline\hline
\multicolumn{6}{|c|}{OLYMPUS Experiment}\\
\hline
\multicolumn{2}{|c|}{ } 
& (a)  & (b) & (c) & (d)   \\
\noalign{\smallskip}\hline
\multicolumn{2}{|c|}  {$ <R_{2\gamma}>$ }& 0.992$\pm$  0.001 &  0.997$\pm$ 0.001 & 0.994 $\pm$  0.001 & 0.994$\pm$ 0.002 \\
 \multicolumn{2}{|c|} {$\chi ^2/N$(1)   }       & 23.2/18=1.29          &11.5/18=0.64         & 15.0/18=0.85           &        8.64/18= 0.48 \\
\multicolumn{2}{|c|}  {All Data $ <R_{2\gamma}>$ }          & 0.999$\pm$  0.001 &  0.999$\pm$ 0.001 & 1.000 $\pm$  0.001 & 1.000$\pm$ 0.002 \\
\multicolumn{2}{|c|}  {All Data  $\chi ^2/N$(1)  }                & 69.3/35=1.98             &56.7/35=1.62         & 60.5/35=1.73      & 53.6/35=1.53 \\
 \noalign{\smallskip}\hline
\multirow{2}{*}{$R_{2\gamma}=a_0+a_1 \epsilon$}    
  &$a_0$    &1.002$\pm$  0.014   & 1.025$\pm$  0.014  &  1.001$\pm$ 0.014 & 1.0167 $\pm$  0.0145  \\
  &$a_1$    & -0.012$\pm$  0.017 &  -0.035$\pm$ 0.017 & -0.007 $\pm$  0.017 & -0.024 $\pm$  0.0174  \\
 % \multicolumn{2}{|c|}{    $\chi ^2/N$    } &38.6/34= 1.13& 5.44/18= 0.3   &7.76/18 =0.97     & 3.08/2 =1.5            \\
 \cline{2-6}
\multicolumn{2}{|c|}{  $\chi ^2/N$ }    &5.44/18= 0.30& 5.04/18= 0.28   &7.76/18 =0.43     & 4.49/18 =0.25            \\
    \hline
 \multirow{2}{*}{$R_{2\gamma}=b_0+ b_1 Q^2$ }       
    &$b_0$& 0.990$\pm $0.0046 &0.989 $\pm$  0.0045 &  0.9935$\pm$ 0.0047 & 0.992 $\pm$  0.055  \\  
   &$b_1$&  0.0019$\pm$  0.0046 &0.0082$\pm$  0.0046 &  0.0013$\pm$ 0.0046 & 0.0055 $\pm$  0.0046  \\
 \cline{2-6}
\multicolumn{2}{|c|}{ $\chi ^2/N$ }     &5.74/18=  0.32   &5.95/18=0.33       & 7.87/18=0.44   & 4.98/18=0.61 \\
    \hline
\end{tabular}
\caption{Experimental data for the four OLYMPUS analysis . The weighted average of the ratio $R_ {2\gamma}$   to be compared to 1 for no $2\gamma e $ contribution. The compatibility with a constant $R_ {2\gamma}=1$ is  indicated  by  $\chi ^2$(1) per number of point; the weighted average of the ratio $R_ {2\gamma}$ for all data slightly changes when the four different assets for soft radiative correction are applied. The results of a linear fit in $\epsilon$ and $Q^2$ are also shown.}
\label{Table:Table2}       % Give a unique label
\end{table}
%\end{document}

\begin{table}
\begin{tabular}{|c|c|c|c|c|c|}
\hline\hline
\multicolumn{6}{|c|}{OLYMPUS Experiment}\\
\hline
    \multicolumn{6}{|c|}{Difference Theory-Experiment}\\
\noalign{\smallskip}\hline
\multicolumn{2}{|c|}{ }  & (a)  & (b) & (c) & (d)   \\
\noalign{\smallskip}\hline
\multicolumn{2}{|c|}{$<R_{Th}-R_{2\gamma}>_{MTj}$}    & 0.009$\pm$ 0.002 &  0.003 $\pm$ 0.001 & 0.002 $\pm$ 0.001 & 0.001$\pm$ 0.001 \\
   \multicolumn{2}{|c|}{    $\chi ^2/N$(0)    }           &  31.5/18=1.75     & 10.8/18=0.65    & 21.9/18= 1.22     & 10.7/18=0.59        \\
\noalign{\smallskip}\hline
\multirow{2}{*}{$<R_{Th}-R_{2\gamma}>_{MTj}=c_0+c_1 \epsilon$}    
  &$c_0$    &0.021$\pm$  0.014   & -0.003$\pm$  0.014  &  1.001$\pm$ 0.014 & 1.0167 $\pm$  0.0145  \\
  &$c_1$    & -0.013$\pm$  0.017 &  0.0099 $\pm$ 0.017 & -0.018$\pm$ 0.018 & -0.024 $\pm$  0.0174  \\
 % \multicolumn{2}{|c|}{    $\chi ^2/N$    } &38.6/34= 1.13& 5.44/18= 0.3   &7.76/18 =0.97     & 3.08/2 =1.5            \\
 \cline{2-6}
\multicolumn{2}{|c|}{  $\chi ^2/N$ }    &4.49/18= 0.25& 4.27/18= 0.24   &7.29/18 =0.41     & 3.88/18 =0.22           \\
    \hline
 \multirow{2}{*}{$R_{2\gamma}=d_0+ d_1 Q^2$ }       
    &$b_0$& 0.006$\pm $0.0046 &0.0068 $\pm$ 0.0046   &  0.0024 $\pm$ 0.0046 & 0.0041 $\pm$  0.0046  \\  
   &$b_1$&  0.0043$\pm$  0.0046 & -0.0018$\pm$0.0046   &  0.0005$\pm$ 0.0046 & 0.0009 $\pm$  0.0046  \\
 \cline{2-6}
\multicolumn{2}{|c|}{ $\chi ^2/N$ }     &4.19/18=  0.23   &4.42/18=0.25    & 7.17/18=0.40   & 3.86/18=0.21 \\
    \hline
\noalign{\smallskip}\hline
\multirow{2}{*}{$<R_{Th}-R_{2\gamma}>_{MTs}=e_0+e_1 \epsilon$}    
  &$e_0$    &0.012$\pm$  0.014   & -0.013$\pm$  0.014  &  -0.003$\pm$ 0.014 & 0.0068 $\pm$  0.0046  \\
  &$e_1$    & -0.003$\pm$  0.017 &  -0.0073 $\pm$ 0.017 & 0.009$\pm$ &  -0.0017 $\pm$ 0.005  \\
 \cline{2-6}
\multicolumn{2}{|c|}{  $\chi ^2/N$ }    &5.20/18= 0.25& 4.84/18= 0.24   &7.63/18 =0.41     & 4.33/18 =0.22           \\
    \hline
 \multirow{2}{*}{$R_{2\gamma}=f_0+ f_1 Q^2$ }       
    &$f_0$& 0.0082$\pm $0.0046 &0.0095 $\pm$ 0.0046   &  0.0051 $\pm$ 0.0046 & 0.0068 $\pm$  0.0046  \\  
   &$f_1$&  0.0018$\pm$  0.0046 & -0.0044$\pm$0.0046   &  0.0024$\pm$ 0.0046 & -0.0017 $\pm$  0.0046  \\
 \cline{2-6}
\multicolumn{2}{|c|}{ $\chi ^2/N$ }     &5.08/18=  0.23   &5.29/18=0.25    & 7.52/18=0.40   & 4.48/18=0.21 \\
    \hline
\end{tabular}
\caption{Difference between the data and the calculation, for the four OLYMPUS analysis: weighted average of the ratio $R_ {2\gamma}$   to be compared to 1 for no $2\gamma e $ contribution. The compatibility with a constant $R_ {2\gamma}=1$ is  indicated  by  $\chi ^2$(1) per number of point; the weighted average of the ratio $R_ {2\gamma}$ for all data slightly changes when the four different assets for soft radiative correction are applied (3rd line). The difference between the data and the calculation of Ref. \cite{Kuraev:2007dn} is reported after subtraction of the soft corrections  from  \cite{Maximon:2000hm}  and from Ref. \cite{Mo:1968cg}. The  compatibility of such difference with a zero constant is seen from $\chi ^2$(0). The linear fit in$\epsilon$ and $Q^2$ is also shown.}
\label{Table:Table2a}       % Give a unique label
\end{table}

\section{Conclusions}

This paper compares the calculation from Ref. \cite{Kuraev:2007dn} to the recent and precise data on elastic scattering of electrons and  positrons on protons.  The ratio contains the information on charge-odd contributions to the cross section and to an eventual contribution of two photon exchange. The raw ratio, when published data are available, is in agreement with the odd contribution deriving mainly from the interference between initial and final state emission.

This work completes (and is consistent with) the analysis published in Ref. \cite{TomasiGustafsson:2009pw}, that reviewed the data of interest for the problem discussed here that were present before the recent experiments.

We stress that the extraction of the 'hard' two photon contribution is somewhat ambiguous as depends on the model used for the implemented radiative corrections, the main problem being the subtraction of the infrared divergent part.  If this subtraction may be straightforward in the calculation, it definitely originates differences in the Monte Carlo implementation. Even if the same model for the radiative correction  is used in the different experiments, what is not the case, the data are corrected with dedicated Monte Carlo, implemented for the specific experiment. The numerical approximations and cuts, that depend on the relevant kinematical variables, are handled differently by the different collaborations. Moreover radiative corrections are implemented together with acceptance and efficiency corrections, that are specific to the individual set-up, making impossible a quantitatively precise comparison. 

Nevertheless, we would like to stress that $R_{2\gamma}$, as 
measured in the experiment, is the ratio of even and odd corrections and all 
corrections due to the efficiency of the detector are factorized in the 
Born-like terms and cancel in the ratio. So the measured 
ratio  must be independent of the different experimental setups, at least at the 
leading terms of corrections.

We minimized this effect by subtracting the applied radiative corrections and replacing by the calculation from Ref. \cite{Kuraev:2007dn}.  For the VEPP and CLAS experiments, when the raw data are available before implementing radiative corrections, we evaluated the effect of the deconvolution between the soft and hard terms. 

The procedure of subtracting two models of radiative corrections, in the same kinematical conditions enhances the model-dependent difference. A similar procedure was validated in previous works, for example \cite{Pacetti:2016tqi,Bystritskiy:2007hw} for a reanalysis of the Andivahis elastic $ep$ scattering data. Not having  data and Monte Carlo in hands, we can still compare the effect of different calculations: they depend only on one parameter, the elasticity cut, $\Delta E$, that can be reasonably estimated.
We stress that the final result on $R_{2\gamma}$ remains independent from $\Delta E$. 

The conclusions of the recent experimental papers, are far from being definite statements. The common issue  is that measurements at large $Q^2$ are necessary. All existing model dependent and independent estimations predict a small effect at low $Q^2$. In absence of specific nuclear effects, QED predicts a  hard \TPE contribution of the percent level with respect to the main (Born) contribution, with mild $\epsilon$ and $Q^2$ dependence \cite{Kuraev:2009hj}, what is indeed seen. Other explanations to fully explain the discrepancy between the unpolarized and polarized form factor ratio experiments are likely to be preferred.

In the OLYMPUS paper it is clearly stated that { \it  We do not agree with the 
conclusions of earlier papers [25,26]. The data shown in Fig. 3 
clearly favors a smaller $R_{2y}$.....To clarify the
situation, the size of TPE at large Q2 has to be determined
in future measurements. }

The VEPP publication concludes  {\it on a significant two-photon exchange effect} , nuanced by a discussion on the used normalization and by the statement that the data are {\it in moderate agreement with several TPE (Two Photon Exchange)  predictions explaining the form factor discrepancy at high $Q^2$}......'

In the CLAS publications  one finds the following statement in the abstract {\it Our results..demonstrate a nonzero contribution from TPE effects and are in excellent agreement with the calculations that include TPE effects and largely reconcile the form-factor discrepancy up to $Q^2\simeq 2$ GeV$^2$ } somehow nuanced in the Conclusions
"{\it  experiments .. to extend the measurements to $Q^2>$ 3 GeV$^2$ ...are needed before one can definitely state that TPE effects are the reason of the discrepancy}" . 

We do not enter here in the comparison and the virtues of the model dependent \TPE calculations. Let us note that, if a qualitative agreement is found on reproducing the difference between polarized and unpolarized FF ratio, the agreement disappear when compared to another observable, the $\epsilon$ dependence of $P_L/P_t$ \cite{Meziane:2010xc}.
Does  the discrepancy between the unpolarized and polarized form factor ratio experiments really exist?
Following the recent work \cite{Pacetti:2015iqa} a problem of renormalization of the low $\epsilon$ data in the previous Rosenbluth analysis, in particular in  Ref. \cite{Andivahis:1994rq}, was pointed out. Then,  the discrepancy remains only for the data from Ref. \cite{Qattan:2004ht}, for which, however, the applied radiative corrections are not known, and a 100\% correlation of the parameters was illustrated in Ref. \cite{TomasiGustafsson:2006pa}.

We confirm the conclusions of that paper of no evident enhancement of the \TPE contribution in the considered data. Our works support alternative explanations to the issue of the form factor discrepancy, if any.
%%%%%%%%%%%%%%%%%%%%%%%%
\subsection{Acknowledgments}
%%%%%%%%%%%%%%%%%%%%%%%%
We acknowledge V. Fadin and to D. Nikolenko for interest in this work and useful discussions. Thanks are due to B. Raue for clarifying issues concerning the CLAS results and providing data in a tabulated form.
%\bibliography{Biblio}

\begin{thebibliography}{39}
\expandafter\ifx\csname natexlab\endcsname\relax\def\natexlab#1{#1}\fi
\expandafter\ifx\csname bibnamefont\endcsname\relax
  \def\bibnamefont#1{#1}\fi
\expandafter\ifx\csname bibfnamefont\endcsname\relax
  \def\bibfnamefont#1{#1}\fi
\expandafter\ifx\csname citenamefont\endcsname\relax
  \def\citenamefont#1{#1}\fi
\expandafter\ifx\csname url\endcsname\relax
  \def\url#1{\texttt{#1}}\fi
\expandafter\ifx\csname urlprefix\endcsname\relax\def\urlprefix{URL }\fi
\providecommand{\bibinfo}[2]{#2}
\providecommand{\eprint}[2][]{\url{#2}}

\bibitem[{\citenamefont{Pacetti et~al.}(2015)\citenamefont{Pacetti,
  Baldini~Ferroli, and Tomasi-Gustafsson}}]{Pacetti:2015iqa}
\bibinfo{author}{\bibfnamefont{S.}~\bibnamefont{Pacetti}},
  \bibinfo{author}{\bibfnamefont{R.}~\bibnamefont{Baldini~Ferroli}},
  \bibnamefont{and}
  \bibinfo{author}{\bibfnamefont{E.}~\bibnamefont{Tomasi-Gustafsson}},
  \bibinfo{journal}{Phys. Rep.} \textbf{\bibinfo{volume}{550-551}},
  \bibinfo{pages}{1} (\bibinfo{year}{2015}).

\bibitem[{\citenamefont{Akhiezer and Rekalo}(1968)}]{Akhiezer:1968ek}
\bibinfo{author}{\bibfnamefont{A.}~\bibnamefont{Akhiezer}} \bibnamefont{and}
  \bibinfo{author}{\bibfnamefont{M.}~\bibnamefont{Rekalo}},
  \bibinfo{journal}{Sov. Phys. Dokl.} \textbf{\bibinfo{volume}{13}},
  \bibinfo{pages}{572} (\bibinfo{year}{1968}).

\bibitem[{\citenamefont{Akhiezer and Rekalo}(1974)}]{Akhiezer:1974em}
\bibinfo{author}{\bibfnamefont{A.}~\bibnamefont{Akhiezer}} \bibnamefont{and}
  \bibinfo{author}{\bibfnamefont{M.}~\bibnamefont{Rekalo}},
  \bibinfo{journal}{Sov. J. Part. Nucl.} \textbf{\bibinfo{volume}{4}},
  \bibinfo{pages}{277} (\bibinfo{year}{1974}).

\bibitem[{\citenamefont{Rosenbluth}(1950)}]{Rosenbluth:1950yq}
\bibinfo{author}{\bibfnamefont{M.}~\bibnamefont{Rosenbluth}},
  \bibinfo{journal}{Phys. Rev.} \textbf{\bibinfo{volume}{79}},
  \bibinfo{pages}{615} (\bibinfo{year}{1950}).

\bibitem[{\citenamefont{Puckett et~al.}(2017)}]{Puckett:2017flj}
\bibinfo{author}{\bibfnamefont{A.~J.~R.} \bibnamefont{Puckett}}
  \bibnamefont{et~al.}, \bibinfo{journal}{Phys. Rev.}
  \textbf{\bibinfo{volume}{C96}}, \bibinfo{pages}{055203}
  (\bibinfo{year}{2017}).

\bibitem[{\citenamefont{Bystritskiy et~al.}(2007)\citenamefont{Bystritskiy,
  Kuraev, and Tomasi-Gustafsson}}]{Bystritskiy:2007hw}
\bibinfo{author}{\bibfnamefont{Y.}~\bibnamefont{Bystritskiy}},
  \bibinfo{author}{\bibfnamefont{E.}~\bibnamefont{Kuraev}}, \bibnamefont{and}
  \bibinfo{author}{\bibfnamefont{E.}~\bibnamefont{Tomasi-Gustafsson}},
  \bibinfo{journal}{Phys. Rev.} \textbf{\bibinfo{volume}{C75}},
  \bibinfo{pages}{015207} (\bibinfo{year}{2007}).

\bibitem[{\citenamefont{Gramolin and Nikolenko}(2016)}]{Gramolin:2016hjt}
\bibinfo{author}{\bibfnamefont{A.~V.} \bibnamefont{Gramolin}} \bibnamefont{and}
  \bibinfo{author}{\bibfnamefont{D.~M.} \bibnamefont{Nikolenko}},
  \bibinfo{journal}{Phys. Rev.} \textbf{\bibinfo{volume}{C93}},
  \bibinfo{pages}{055201} (\bibinfo{year}{2016}).

\bibitem[{\citenamefont{Gerasimov and Fadin}(2015)}]{Gerasimov:2015aoa}
\bibinfo{author}{\bibfnamefont{R.~E.} \bibnamefont{Gerasimov}}
  \bibnamefont{and} \bibinfo{author}{\bibfnamefont{V.~S.} \bibnamefont{Fadin}},
  \bibinfo{journal}{Phys. Atom. Nucl.} \textbf{\bibinfo{volume}{78}},
  \bibinfo{pages}{69} (\bibinfo{year}{2015}), \bibinfo{note}{[Yad.
  Fiz.78,no.1-2,73(2015)]}.

\bibitem[{\citenamefont{Tomasi-Gustafsson}(2007)}]{TomasiGustafsson:2006pa}
\bibinfo{author}{\bibfnamefont{E.}~\bibnamefont{Tomasi-Gustafsson}},
  \bibinfo{journal}{Phys. Part. Nucl. Lett.} \textbf{\bibinfo{volume}{4}},
  \bibinfo{pages}{281} (\bibinfo{year}{2007}).

\bibitem[{\citenamefont{Pacetti and Tomasi-Gustafsson}(2016)}]{Pacetti:2016tqi}
\bibinfo{author}{\bibfnamefont{S.}~\bibnamefont{Pacetti}} \bibnamefont{and}
  \bibinfo{author}{\bibfnamefont{E.}~\bibnamefont{Tomasi-Gustafsson}},
  \bibinfo{journal}{Phys. Rev.} \textbf{\bibinfo{volume}{C94}},
  \bibinfo{pages}{055202} (\bibinfo{year}{2016}).

\bibitem[{\citenamefont{Arrington}(2003)}]{Arrington:2003df}
\bibinfo{author}{\bibfnamefont{J.}~\bibnamefont{Arrington}},
  \bibinfo{journal}{Phys. Rev.} \textbf{\bibinfo{volume}{C68}},
  \bibinfo{pages}{034325} (\bibinfo{year}{2003}), \eprint{nucl-ex/0305009}.

\bibitem[{\citenamefont{De~Rujula et~al.}(1971)\citenamefont{De~Rujula, Kaplan,
  and De~Rafael}}]{DeRujula:1972te}
\bibinfo{author}{\bibfnamefont{A.}~\bibnamefont{De~Rujula}},
  \bibinfo{author}{\bibfnamefont{J.~M.} \bibnamefont{Kaplan}},
  \bibnamefont{and}
  \bibinfo{author}{\bibfnamefont{E.}~\bibnamefont{De~Rafael}},
  \bibinfo{journal}{Nucl. Phys.} \textbf{\bibinfo{volume}{B35}},
  \bibinfo{pages}{365} (\bibinfo{year}{1971}).

\bibitem[{\citenamefont{De~Rujula et~al.}(1973)\citenamefont{De~Rujula, Kaplan,
  and De~Rafael}}]{DeRujula:1973pr}
\bibinfo{author}{\bibfnamefont{A.}~\bibnamefont{De~Rujula}},
  \bibinfo{author}{\bibfnamefont{J.~M.} \bibnamefont{Kaplan}},
  \bibnamefont{and}
  \bibinfo{author}{\bibfnamefont{E.}~\bibnamefont{De~Rafael}},
  \bibinfo{journal}{Nucl. Phys.} \textbf{\bibinfo{volume}{B53}},
  \bibinfo{pages}{545} (\bibinfo{year}{1973}).

\bibitem[{\citenamefont{Afanasev et~al.}(2017)\citenamefont{Afanasev, Blunden,
  Hasell, and Raue}}]{Afanasev:2017gsk}
\bibinfo{author}{\bibfnamefont{A.}~\bibnamefont{Afanasev}},
  \bibinfo{author}{\bibfnamefont{P.~G.} \bibnamefont{Blunden}},
  \bibinfo{author}{\bibfnamefont{D.}~\bibnamefont{Hasell}}, \bibnamefont{and}
  \bibinfo{author}{\bibfnamefont{B.~A.} \bibnamefont{Raue}},
  \bibinfo{journal}{Prog. Part. Nucl. Phys.} \textbf{\bibinfo{volume}{95}},
  \bibinfo{pages}{245} (\bibinfo{year}{2017}).

\bibitem[{\citenamefont{Gunion and Stodolsky}(1973)}]{Gunion:1972bj}
\bibinfo{author}{\bibfnamefont{J.}~\bibnamefont{Gunion}} \bibnamefont{and}
  \bibinfo{author}{\bibfnamefont{L.}~\bibnamefont{Stodolsky}},
  \bibinfo{journal}{Phys. Rev. Lett.} \textbf{\bibinfo{volume}{30}},
  \bibinfo{pages}{345} (\bibinfo{year}{1973}).

\bibitem[{\citenamefont{Boitsov et~al.}(1973)\citenamefont{Boitsov, Kondratyuk,
  and Kopeliovich}}]{Boitsov:1972if}
\bibinfo{author}{\bibfnamefont{V.}~\bibnamefont{Boitsov}},
  \bibinfo{author}{\bibfnamefont{L.}~\bibnamefont{Kondratyuk}},
  \bibnamefont{and}
  \bibinfo{author}{\bibfnamefont{V.}~\bibnamefont{Kopeliovich}},
  \bibinfo{journal}{Sov. J. Nucl. Phys.} \textbf{\bibinfo{volume}{16}},
  \bibinfo{pages}{287} (\bibinfo{year}{1973}).

\bibitem[{\citenamefont{Franco}(1973)}]{Franco:1973uq}
\bibinfo{author}{\bibfnamefont{V.}~\bibnamefont{Franco}},
  \bibinfo{journal}{Phys. Rev.} \textbf{\bibinfo{volume}{D8}},
  \bibinfo{pages}{826} (\bibinfo{year}{1973}).

\bibitem[{\citenamefont{Tomasi-Gustafsson
  et~al.}(2013)\citenamefont{Tomasi-Gustafsson, Osipenko, Kuraev, and
  Bystritsky}}]{TomasiGustafsson:2009pw}
\bibinfo{author}{\bibfnamefont{E.}~\bibnamefont{Tomasi-Gustafsson}},
  \bibinfo{author}{\bibfnamefont{M.}~\bibnamefont{Osipenko}},
  \bibinfo{author}{\bibfnamefont{E.}~\bibnamefont{Kuraev}}, \bibnamefont{and}
  \bibinfo{author}{\bibfnamefont{Y.}~\bibnamefont{Bystritsky}},
  \bibinfo{journal}{Phys. Atom. Nucl.} \textbf{\bibinfo{volume}{76}},
  \bibinfo{pages}{937} (\bibinfo{year}{2013}).

\bibitem[{\citenamefont{Rekalo and
  Tomasi-Gustafsson}(2004{\natexlab{a}})}]{Rekalo:2003xa}
\bibinfo{author}{\bibfnamefont{M.~P.} \bibnamefont{Rekalo}} \bibnamefont{and}
  \bibinfo{author}{\bibfnamefont{E.}~\bibnamefont{Tomasi-Gustafsson}},
  \bibinfo{journal}{Eur. Phys. J.} \textbf{\bibinfo{volume}{A22}},
  \bibinfo{pages}{331} (\bibinfo{year}{2004}{\natexlab{a}}).

\bibitem[{\citenamefont{Rekalo and
  Tomasi-Gustafsson}(2004{\natexlab{b}})}]{Rekalo:2003km}
\bibinfo{author}{\bibfnamefont{M.}~\bibnamefont{Rekalo}} \bibnamefont{and}
  \bibinfo{author}{\bibfnamefont{E.}~\bibnamefont{Tomasi-Gustafsson}},
  \bibinfo{journal}{Nucl. Phys.} \textbf{\bibinfo{volume}{A740}},
  \bibinfo{pages}{271} (\bibinfo{year}{2004}{\natexlab{b}}).

\bibitem[{\citenamefont{Rekalo and
  Tomasi-Gustafsson}(2004{\natexlab{c}})}]{Rekalo:2004wa}
\bibinfo{author}{\bibfnamefont{M.}~\bibnamefont{Rekalo}} \bibnamefont{and}
  \bibinfo{author}{\bibfnamefont{E.}~\bibnamefont{Tomasi-Gustafsson}},
  \bibinfo{journal}{Nucl. Phys.} \textbf{\bibinfo{volume}{A742}},
  \bibinfo{pages}{322} (\bibinfo{year}{2004}{\natexlab{c}}).

\bibitem[{\citenamefont{Rekalo et~al.}(1999)\citenamefont{Rekalo,
  Tomasi-Gustafsson, and Prout}}]{Rekalo:1999mt}
\bibinfo{author}{\bibfnamefont{M.~P.} \bibnamefont{Rekalo}},
  \bibinfo{author}{\bibfnamefont{E.}~\bibnamefont{Tomasi-Gustafsson}},
  \bibnamefont{and} \bibinfo{author}{\bibfnamefont{D.}~\bibnamefont{Prout}},
  \bibinfo{journal}{Phys. Rev.} \textbf{\bibinfo{volume}{C60}},
  \bibinfo{pages}{042202} (\bibinfo{year}{1999}).

\bibitem[{\citenamefont{Kuraev et~al.}(2006)\citenamefont{Kuraev, Bytev,
  Bystritskiy, and Tomasi-Gustafsson}}]{Kuraev:2006ys}
\bibinfo{author}{\bibfnamefont{E.}~\bibnamefont{Kuraev}},
  \bibinfo{author}{\bibfnamefont{V.}~\bibnamefont{Bytev}},
  \bibinfo{author}{\bibfnamefont{Y.}~\bibnamefont{Bystritskiy}},
  \bibnamefont{and}
  \bibinfo{author}{\bibfnamefont{E.}~\bibnamefont{Tomasi-Gustafsson}},
  \bibinfo{journal}{Phys. Rev.} \textbf{\bibinfo{volume}{D74}},
  \bibinfo{pages}{013003} (\bibinfo{year}{2006}).

\bibitem[{\citenamefont{Kuraev et~al.}(2008)\citenamefont{Kuraev, Bytev,
  Bakmaev, and Tomasi-Gustafsson}}]{Kuraev:2007dn}
\bibinfo{author}{\bibfnamefont{E.}~\bibnamefont{Kuraev}},
  \bibinfo{author}{\bibfnamefont{V.}~\bibnamefont{Bytev}},
  \bibinfo{author}{\bibfnamefont{S.}~\bibnamefont{Bakmaev}}, \bibnamefont{and}
  \bibinfo{author}{\bibfnamefont{E.}~\bibnamefont{Tomasi-Gustafsson}},
  \bibinfo{journal}{Phys. Rev.} \textbf{\bibinfo{volume}{C78}},
  \bibinfo{pages}{015205} (\bibinfo{year}{2008}).

\bibitem[{\citenamefont{Tomalak et~al.}(2017)\citenamefont{Tomalak, Pasquini,
  and Vanderhaeghen}}]{Tomalak:2017shs}
\bibinfo{author}{\bibfnamefont{O.}~\bibnamefont{Tomalak}},
  \bibinfo{author}{\bibfnamefont{B.}~\bibnamefont{Pasquini}}, \bibnamefont{and}
  \bibinfo{author}{\bibfnamefont{M.}~\bibnamefont{Vanderhaeghen}},
  \bibinfo{journal}{Phys. Rev.} \textbf{\bibinfo{volume}{D96}},
  \bibinfo{pages}{096001} (\bibinfo{year}{2017}), \eprint{1708.03303}.

\bibitem[{\citenamefont{Maximon and Tjon}(2000)}]{Maximon:2000hm}
\bibinfo{author}{\bibfnamefont{L.}~\bibnamefont{Maximon}} \bibnamefont{and}
  \bibinfo{author}{\bibfnamefont{J.}~\bibnamefont{Tjon}},
  \bibinfo{journal}{Phys. Rev.} \textbf{\bibinfo{volume}{C62}},
  \bibinfo{pages}{054320} (\bibinfo{year}{2000}).

\bibitem[{\citenamefont{Mo and Tsai}(1969)}]{Mo:1968cg}
\bibinfo{author}{\bibfnamefont{L.~W.} \bibnamefont{Mo}} \bibnamefont{and}
  \bibinfo{author}{\bibfnamefont{Y.-S.} \bibnamefont{Tsai}},
  \bibinfo{journal}{Rev. Mod. Phys.} \textbf{\bibinfo{volume}{41}},
  \bibinfo{pages}{205} (\bibinfo{year}{1969}).

\bibitem[{\citenamefont{Ent et~al.}(2001)\citenamefont{Ent, Filippone, Makins,
  Milner, O'Neill, and Wasson}}]{Ent:2001hm}
\bibinfo{author}{\bibfnamefont{R.}~\bibnamefont{Ent}},
  \bibinfo{author}{\bibfnamefont{B.~W.} \bibnamefont{Filippone}},
  \bibinfo{author}{\bibfnamefont{N.~C.~R.} \bibnamefont{Makins}},
  \bibinfo{author}{\bibfnamefont{R.~G.} \bibnamefont{Milner}},
  \bibinfo{author}{\bibfnamefont{T.~G.} \bibnamefont{O'Neill}},
  \bibnamefont{and} \bibinfo{author}{\bibfnamefont{D.~A.}
  \bibnamefont{Wasson}}, \bibinfo{journal}{Phys. Rev.}
  \textbf{\bibinfo{volume}{C64}}, \bibinfo{pages}{054610}
  (\bibinfo{year}{2001}).

\bibitem[{\citenamefont{Gramolin et~al.}(2014)\citenamefont{Gramolin, Fadin,
  Feldman, Gerasimov, Nikolenko, Rachek, and Toporkov}}]{Gramolin:2014pva}
\bibinfo{author}{\bibfnamefont{A.~V.} \bibnamefont{Gramolin}},
  \bibinfo{author}{\bibfnamefont{V.~S.} \bibnamefont{Fadin}},
  \bibinfo{author}{\bibfnamefont{A.~L.} \bibnamefont{Feldman}},
  \bibinfo{author}{\bibfnamefont{R.~E.} \bibnamefont{Gerasimov}},
  \bibinfo{author}{\bibfnamefont{D.~M.} \bibnamefont{Nikolenko}},
  \bibinfo{author}{\bibfnamefont{I.~A.} \bibnamefont{Rachek}},
  \bibnamefont{and} \bibinfo{author}{\bibfnamefont{D.~K.}
  \bibnamefont{Toporkov}}, \bibinfo{journal}{J. Phys.}
  \textbf{\bibinfo{volume}{G41}}, \bibinfo{pages}{115001}
  (\bibinfo{year}{2014}).

\bibitem[{\citenamefont{Meister and Yennie}(1963)}]{Meister:1963zz}
\bibinfo{author}{\bibfnamefont{N.}~\bibnamefont{Meister}} \bibnamefont{and}
  \bibinfo{author}{\bibfnamefont{D.}~\bibnamefont{Yennie}},
  \bibinfo{journal}{Phys. Rev.} \textbf{\bibinfo{volume}{130}},
  \bibinfo{pages}{1210} (\bibinfo{year}{1963}).

\bibitem[{\citenamefont{Tsai}(1961)}]{Tsai:1961zz}
\bibinfo{author}{\bibfnamefont{Y.-S.} \bibnamefont{Tsai}},
  \bibinfo{journal}{Phys. Rev.} \textbf{\bibinfo{volume}{122}},
  \bibinfo{pages}{1898} (\bibinfo{year}{1961}).

\bibitem[{\citenamefont{Rachek et~al.}(2015)}]{Rachek:2014fam}
\bibinfo{author}{\bibfnamefont{I.~A.} \bibnamefont{Rachek}}
  \bibnamefont{et~al.}, \bibinfo{journal}{Phys. Rev. Lett.}
  \textbf{\bibinfo{volume}{114}}, \bibinfo{pages}{062005}
  (\bibinfo{year}{2015}).

\bibitem[{\citenamefont{Henderson et~al.}(2017)}]{Henderson:2016dea}
\bibinfo{author}{\bibfnamefont{B.~S.} \bibnamefont{Henderson}}
  \bibnamefont{et~al.} (\bibinfo{collaboration}{OLYMPUS Collaboration}),
  \bibinfo{journal}{Phys. Rev. Lett.} \textbf{\bibinfo{volume}{118}},
  \bibinfo{pages}{092501} (\bibinfo{year}{2017}).

\bibitem[{\citenamefont{Rimal et~al.}(2017)}]{Rimal:2016toz}
\bibinfo{author}{\bibfnamefont{D.}~\bibnamefont{Rimal}} \bibnamefont{et~al.}
  (\bibinfo{collaboration}{CLAS Collaboration}), \bibinfo{journal}{Phys. Rev.}
  \textbf{\bibinfo{volume}{C95}}, \bibinfo{pages}{065201}
  (\bibinfo{year}{2017}).

\bibitem[{\citenamefont{Nikolenko et~al.}(2015)}]{Nikolenko:2015xsa}
\bibinfo{author}{\bibfnamefont{D.~M.} \bibnamefont{Nikolenko}}
  \bibnamefont{et~al.}, \bibinfo{journal}{Phys. Atom. Nucl.}
  \textbf{\bibinfo{volume}{78}}, \bibinfo{pages}{394} (\bibinfo{year}{2015}),
  \bibinfo{note}{[Yad. Fiz.78,423(2015)]}.

\bibitem[{\citenamefont{Kuraev et~al.}(2009)\citenamefont{Kuraev, Shatnev, and
  Tomasi-Gustafsson}}]{Kuraev:2009hj}
\bibinfo{author}{\bibfnamefont{E.~A.} \bibnamefont{Kuraev}},
  \bibinfo{author}{\bibfnamefont{M.}~\bibnamefont{Shatnev}}, \bibnamefont{and}
  \bibinfo{author}{\bibfnamefont{E.}~\bibnamefont{Tomasi-Gustafsson}},
  \bibinfo{journal}{Phys. Rev.} \textbf{\bibinfo{volume}{C80}},
  \bibinfo{pages}{018201} (\bibinfo{year}{2009}).

\bibitem[{\citenamefont{Meziane et~al.}(2011)}]{Meziane:2010xc}
\bibinfo{author}{\bibfnamefont{M.}~\bibnamefont{Meziane}} \bibnamefont{et~al.}
  (\bibinfo{collaboration}{GEp2gamma}), \bibinfo{journal}{Phys. Rev. Lett.}
  \textbf{\bibinfo{volume}{106}}, \bibinfo{pages}{132501}
  (\bibinfo{year}{2011}).

\bibitem[{\citenamefont{Andivahis et~al.}(1994)\citenamefont{Andivahis, Bosted,
  Lung, Stuart, Alster et~al.}}]{Andivahis:1994rq}
\bibinfo{author}{\bibfnamefont{L.}~\bibnamefont{Andivahis}},
  \bibinfo{author}{\bibfnamefont{P.~E.} \bibnamefont{Bosted}},
  \bibinfo{author}{\bibfnamefont{A.}~\bibnamefont{Lung}},
  \bibinfo{author}{\bibfnamefont{L.}~\bibnamefont{Stuart}},
  \bibinfo{author}{\bibfnamefont{J.}~\bibnamefont{Alster}},
  \bibnamefont{et~al.}, \bibinfo{journal}{Phys. Rev.}
  \textbf{\bibinfo{volume}{D50}}, \bibinfo{pages}{5491} (\bibinfo{year}{1994}).

\bibitem[{\citenamefont{Qattan et~al.}(2005)\citenamefont{Qattan, Arrington,
  Segel, Zheng, Aniol et~al.}}]{Qattan:2004ht}
\bibinfo{author}{\bibfnamefont{I.}~\bibnamefont{Qattan}},
  \bibinfo{author}{\bibfnamefont{J.}~\bibnamefont{Arrington}},
  \bibinfo{author}{\bibfnamefont{R.}~\bibnamefont{Segel}},
  \bibinfo{author}{\bibfnamefont{X.}~\bibnamefont{Zheng}},
  \bibinfo{author}{\bibfnamefont{K.}~\bibnamefont{Aniol}},
  \bibnamefont{et~al.}, \bibinfo{journal}{Phys. Rev. Lett.}
  \textbf{\bibinfo{volume}{94}}, \bibinfo{pages}{142301}
  (\bibinfo{year}{2005}).

\end{thebibliography}

\end{document}